\documentstyle[editedvolume]{crckapb} 

\begin{opening}
\title{VLBI imaging of the gravitational lenses\protect\\ 
       B1422+231 and MG\,J0414+0534 } 

\author{E. ROS}
\institute{Max-Planck-Institut f\"ur Radioastronomie, Bonn, Germany}
\author{J.C. GUIRADO}
\author{J.M. MARCAIDE}
\institute{Dep.\ Astronomia i Astrof\'{\i}sica, U.\ Val\`encia, Burjassot, Spain}
\author{M.A. P\'EREZ-TORRES}
\institute{Istituto di Radioastronomia-CNR, Bologna, Italy}
\author{E.E. FALCO}
\institute{Harvard-Smiths.\ Center for Astrophysics, Cambridge MA, US}
\author{J.A. MU\~NOZ}
\institute{Instituto de Astrof\'{\i}sica de Canarias, 
La Laguna, Spain}
\author{A. ALBERDI}
\author{L. LARA}
\institute{Instituto de Astrof\'{\i}sica de Andaluc\'{\i}a/CSIC
Granada, Spain}
\end{opening}

\runningtitle{VLBI imaging of MG\,J0414+0534 and B1422+231}

\begin{document}


\begin{abstract}
We present wide-field images of the quadruple gravitational lenses
B1422+231 and
MG\,J0414+0534 obtained from global Very Long Baseline Interferometry (VLBI)
observations at 8.4\,GHz  on 23 November 1997.
We present also a lens model for MG\,J0414+0534, which 
reproduces the core positions
and flux densities of the VLBI images, combining a singular isothermal 
ellipsoid with external shear,
and a singular isothermal sphere to represent, respectively, the
main lens galaxy and its neighbor, a faint galaxy near one of the
images.
\end{abstract}

\paragraph{\textbf{Introduction.}}

We have embarked on a project of multi-epoch observations 
of the quadruple images of the two gravitationally lensed
objects {B1422+231} and {MG\,J0414+0534} 
to compare possible structural changes in the lensed images 
and measure possible shifts in the relative positions of the sub-images.
Here we present our images from the first epoch
observations on 23 November 1997.


\paragraph{\textbf{Imaging of B1422+231.}}

%
\begin{figure}
\vspace{259pt}
\includegraphics{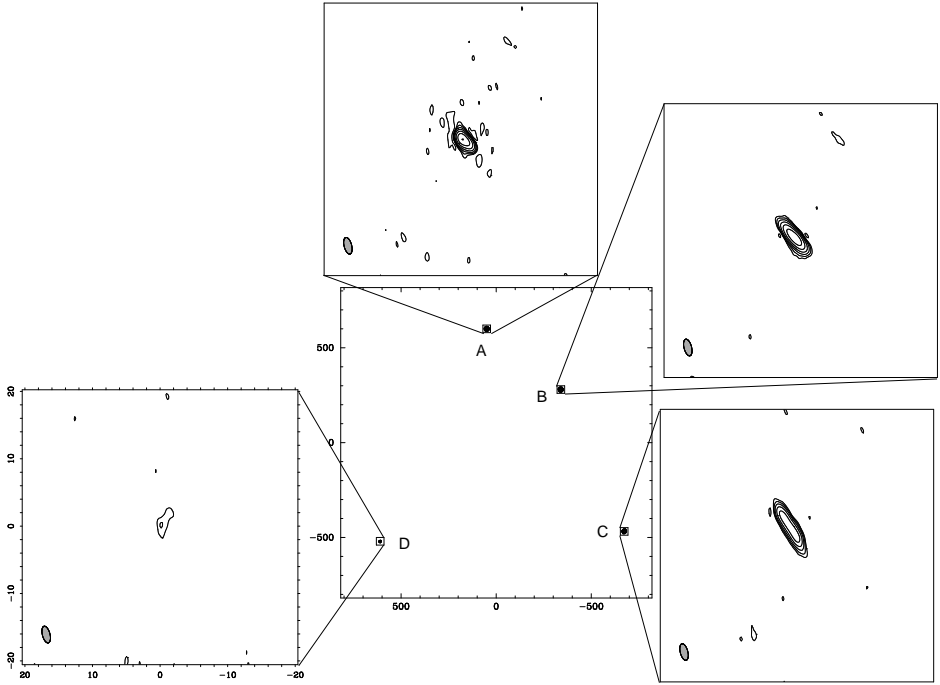}
\caption{Composite of 
our preliminary maps of the four gravitationally lensed images
of {B1422+231} at 8.4\,GHz, labeled as A (top, center), B (top, right),
C (bottom, right), and D (bottom, left) 
with a global VLBI wide-field image (central panel) 
of the radio source.
The axes are relative right ascension and declination, in milliarcseconds. 
The detailed figures are all convolved with a beam of 
1.64$\times$1.16\,mas (position angle (P.A.) 22.43$^\circ$), and
the wide-field image is convolved with a circular 25\,mas beam.
}
\label{fig:maps1422}
\end{figure}

B1422+231 is a quadruple gravitational lens object discovered
by Patnaik et al. (1992).  The maximum image separation between
the four components is 1.3\,arcsec.  The background
radio source is associated with a 15.5\,mag QSO at redshift
$z$=3.62 (Patnaik et al.\ 1992).  The lensing 
object is an elliptical galaxy at 
$z$=0.338
(Tonry 1998).  The lensing system
has been modelled with an elliptical potential and
a shear (see Mao \& Schneider 1998 and references
therein).  The optical images from the Hubble Space Telescope 
(Impey et al.\ 1996)
show the lens object close to the faintest sub-image.
We show in Fig.\ \ref{fig:maps1422}
our VLBI images of B1422+231.
The bright components, A, B, and C are stretched along
the direction joining them, in accordance with (general) lens modeling.
The much weaker image D
shows some elongation towards
the lensing galaxy.
Our images are in good agreement with those 
obtained at 8.4\,GHz 
from Patnaik et al.\ (1999).
The images are consistent with a background compact radio
source without a prominent, extended jet-like structure.

\paragraph{\textbf{Imaging and modeling of MG\,J0414+0534.}}

%
\begin{figure}
\vspace{360pt}
\includegraphics{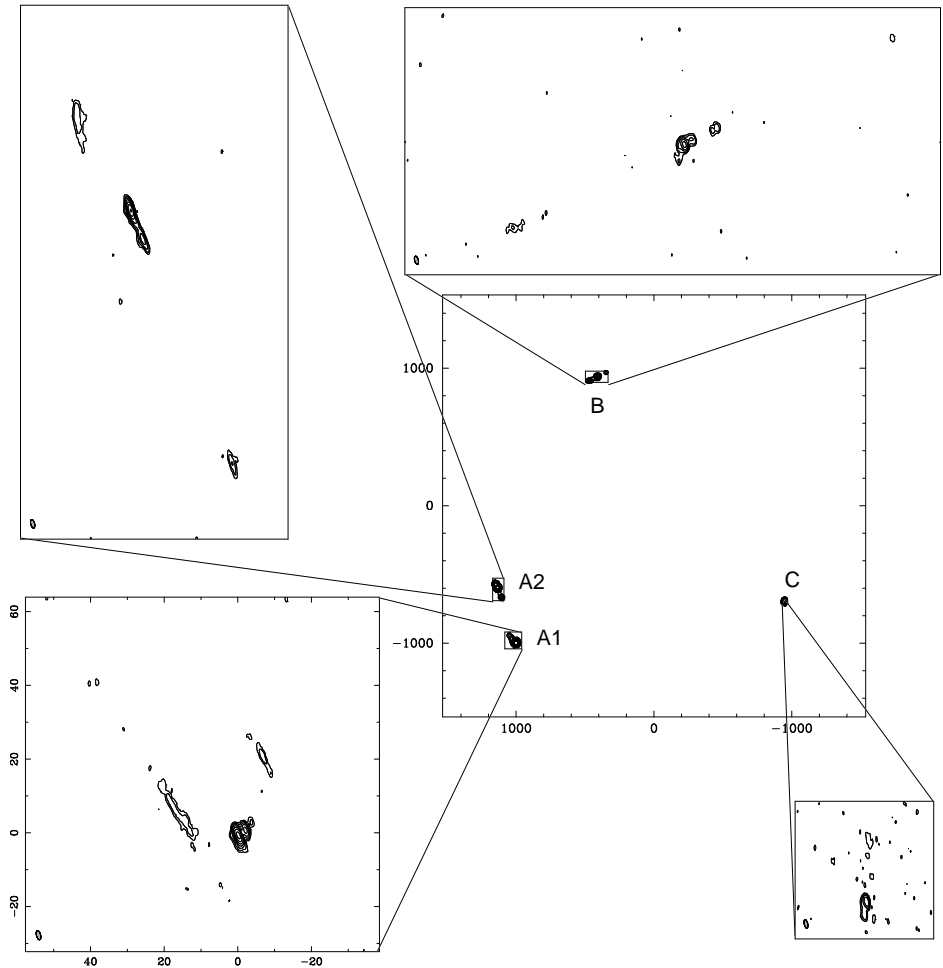}
\caption{Composite of 
our 
images of the four 
gravitationally lensed images of {MG\,J0414+0534}, labeled as A1 (bottom, left), A2 (top,
left), B (top, right), and C (bottom, right) 
at 8.4\,GHz with
a global VLBI wide-field image (central panel) of the radio source.
The axes are relative right ascension and declination, in milliarcseconds. 
The detailed images are convolved with a beam
of 2.55$\times$1.13\,mas (P.A.\ 14.6$^\circ$), and the wide-field
image is convolved with a circular 25\,mas beam.
See also Ros et al.\ (2000).
}
\label{fig:maps0414}
\end{figure}

MG\,J0414+0534 is a quadruple gravitational lens system
discovered by Hewitt et al.\ (1992), with maximum
separations between components of 2.12\,arcsec.
The background object is a 21.2\,mag QSO with
$z$=2.64 (Lawrence et al.\ 1995).  The lensing potential is due
to an
elliptical galaxy at redshift $z$=0.96 (Tonry \& Kochanek 1999) and
its likely neighbor, a galaxy 
$\sim$1.51\,arcsec away
(P.A.\ of $-14^\circ$).
Hubble Space Telescope observations revealed
an optical blue arc joining the three brightest components
(Falco et al.\ 1997).

We present global VLBI images 
of this system
in Fig.\ \ref{fig:maps0414}.
The individual
images exhibit radio
structures extending up to 100\,mas.  The images are
labeled, from the brightest to the weakest as A1, A2, B, and C.
From lens modeling, the structure of B is the most similar
to that of the background radio source,
a double compact structure with a double jet.
A1 shows the most complex structure, with the core-like
region and both jets quite distorted.  A2 is similar to B, but
rotated and mirrored.  For C we only detect emission
at the northern side
of the jet.
The morphology of our images is very similar to that presented
in Trotter et al.\ (2000) at 5\,GHz.

We reproduce successfully the relative positions and
peak of brightness ratios of
the radio cores with a lens model consisting
of a singular isothermal ellipsoid (for the elliptical galaxy
visible at the optical, and placed
at the center between the four radio images) with an external shear,
and a secondary potential given by a singular
isothermal sphere (associated with a small object
close to the west of component B).
The model predicts that B is the leading image, and C lags all
the other images.  The delays from B are of 15.7$\pm$1.3, 16.0$\pm$1.4,
and $66\pm5$\,days for A1, A2, and C, respectively.
In our model, the spherical isothermal ellipse potential
(constrained by the images) is
aligned with the optical images of the lens galaxy.
We plan to refine and study further our lens model by adding the
constraints provided by the lensed jets, using new software 
that
avoids the problem of matching components along these jets, in the
different lensed images. 

We observed these radio sources again
on 8 June 1999, and new observations
are scheduled for late 2000.  
Once these new epochs are analyzed it might be possible to measure
changes in the brightness peak positions
of both radio sources and/or morphological evolution in
the core-like and jet-like regions of
MG\,J0414+0534.

\vspace{-5pt}
\paragraph{Acknowledgements}
This work has
been partially supported by the Spanish DGICYT Grants No.\ PB96-0782
and PB97-1164
and by European Comission's TMR-LSF programme, contract No.\ ERBFMGECT950012.

\end{document}